\documentclass[aps,showpacs,twocolumn]{revtex4}
\usepackage{amssymb}
\usepackage{amsmath}
\usepackage{graphicx}
\usepackage{epsfig}

\begin{document}

\title{Solutions of $\mathcal{PT}$ symmetric tight-binding chain and its
equivalent Hermitian counterpart}
\author{L. Jin and Z. Song}
\email{songtc@nankai.edu.cn}
\affiliation{School of Physics, Nankai University, Tianjin 300071, China}

\begin{abstract}
We study the Non-Hermitian quantum mechanics for the discrete system. This
paper gives an exact analytic single-particle solution for an $N$-site
tight-binding chain with two conjugated imaginary potentials $\pm i\gamma $
at two end sites, which Hamiltonian has parity-time symmetry ($\mathcal{PT}$
symmetry). Based on the Bethe ansatz results, it is found that, in
single-particle subspace, this model is comprised of two phases, an unbroken
symmetry phase with a purely real energy spectrum in the region $\gamma
\prec \gamma _{c}$ and a spontaneously-broken symmetry phase with $N-2$ real
and $2$ imaginary eigenvalues in the region $\gamma \succ \gamma _{c}$. The
behaviors of eigenfunctions and eigenvalues in the vicinity of $\gamma _{c}$
are investigated. It is shown that the boundary of two phases possesses the
characteristics of exceptional point. We also construct the equivalent
Hermitian Hamiltonian\ of the present model in the framework of
metric-operator theory. We find out that the equivalent Hermitian
Hamiltonian can be written as another bipartite lattice model with real
long-range hoppings.
\end{abstract}

\maketitle

\qquad %03.65.-w Quantum mechanics
%11.30.Er Charge conjugation, parity, time reversal, and other discrete symmetries
%71.10.Fd Lattice fermion models (Hubbard model, etc.)
%73.22.Dj Single particle states (ignored)
%73.23.-b Electronic transport in mesoscopic systems (ignored)

\section{Introduction}

Since the discovery that a non-Hermitian Hamiltonian having simultaneous
parity-time ($\mathcal{PT}$) symmetry\ has an entirely real
quantum-mechanical energy spectrum \cite{Bender 98}, there has been an
intense effort to establish a $\mathcal{PT}$-symmetric quantum theory as a
complex extension of the conventional quantum mechanics \cite%
{A.M,A.M36,Jones,Bender 99,Dorey 01,Bender 02}. Although there have been no
experiments to show clearly and definitively that quantum systems defined by
non-Hermitian $\mathcal{PT}$-symmetric Hamiltonians do exist in nature,
many\ models have been proposed to verify theorems and perform numerical and
asymptotic analysis (for a recent review, see \cite{Bender 07} and
references therein). The reality of the spectra is responsible to the $%
\mathcal{PT}$ symmetry: If all the eigenstates of the Hamiltonian are also
eigenstates of $\mathcal{PT}$, then all the eigenvalues are strictly real
and the symmetry is said to be unbroken. Otherwise, the symmetry is said to
be spontaneously broken. In practice, imaginary potential usually appears in
a system to describe physical processes phenomenologically, which have been
investigated under the Non-Hermitian quantum mechanics framework \cite%
{Klaiman1,Znojil,Makris,Musslimani,Bender
08,Jentschura,Fan,A.M38,A.M391,A.M392}. However it is not clear whether
non-Hermitian Hamiltonians describe real physics or are just unrealistic
mathematical objects. It is known, for a diagonalizable Hamiltonian, that
the presence of $\mathcal{PT}$-symmetry implies the pseudo-Hermiticity of
the Hamiltonian \cite{A.M43}, i.e., $\mathcal{PT}$-symmetry is a special
case of pseudo-Hermiticity. A direct way of extracting the physical meaning
of a pseudo-Hermitian Hamiltonian having a real spectrum is to seek for its
Hermitian counterparts \cite{A.M38,A.M391,A.M392}. The metric-operator
theory outlined in \cite{A.M} provides a mapping of such a pseudo-Hermitian
Hamiltonian to an equivalent Hermitian Hamiltonian. The construction of the
latter is usually quite complicated if the Hilbert space of the systems is
infinite-dimensional. It is generally more tractable for lattice systems
with a finite-dimensional single-particle Hilbert space. Under such
circumstances the study of simple models which are exactly solvable and
slight modified version of a practical model, and yet are tractable, is of
particular importance to examine the impact of an imaginary potential on the
feature of eigenvalues and eigenfunctions of a lattice system.

In this paper, we will focus on the non-Hermitian $\mathcal{PT}$-symmetric
quantum theory for a discrete system \cite{Korff, Graefe}, the original form
of which is exploited to describe the solid-state system in condensed matter
physics or coupled quantum devices since the advent of quantum information
theory \cite{Jaksch}. As an illustration, a simple $N$-site tight-binding
chain with the uniform nearest neighbor (NN) hopping integral is concerned.
Such a model is used to describe the Bloch electronic system in condensed
matter physics and now the qubit array relevant to quantum information
applications. The corresponding non-Hermitian $\mathcal{PT}$-symmetric
version is constructed by adding two conjugated imaginary potentials $\pm
i\gamma $\ at the end sites. The objective of this paper aims at the exact
solutions of such a model so as to confirm the non-Hermitian $\mathcal{PT}$%
-symmetric quantum theory for the discrete system. Based on the Bethe ansatz
results, it is found that, in single-particle subspace, this model exhibits
two phases, an unbroken symmetry phase with a purely real energy spectrum
when the potentials are in the region $\gamma \prec \gamma _{c}$ and a
spontaneously-broken symmetry phase with $N-2$ real and $2$ imaginary
eigenvalues when the potentials are in the region $\gamma \succ \gamma _{c}$
Based on the exact solutions, the behaviors of eigenfunctions and
eigenvalues in the vicinity of $\gamma _{c}$\ are investigated. It is shown
that the boundary of two phases possesses the characteristics of exceptional
point: there are two eigenstates coalescing at $\gamma _{c}$\ as well as
with square-root type level repulsion in the vicinity of it. We also
construct the equivalent Hermitian Hamiltonian\ of the present model in the
framework of metric-operator theory. We find out that the equivalent
Hermitian Hamiltonian can be written as another bipartite lattice model with
real long-range hoppings.

This paper is organized as follows, in Sec. II, the model is presented. In
Sec. III, the Bethe ansatz solutions are given. In Sec. IV, we investigate
the characteristics of the critical point $\gamma _{c}$. Sec. V is devoted
to the construction of the equivalent Hermitian Hamiltonian. Sec. VI is the
summary and discussion.

\section{$\mathcal{PT}$-symmetric uniform tight-binding chain}

We consider a simplest discrete system described by a non-Hermitian
Hamiltonian $H$ having the $\mathcal{PT}$ symmetry. It is a tight-binding
chain with uniform nearest neighbor hopping integral and two additional
conjugated imaginary on-site potentials on the two end sites, which can be
written as follows:
\begin{equation}
H=-J\overset{N-1}{\sum_{l=1}}\left( a_{l}^{\dag }a_{l+1}+\text{H.c.}\right)
+i\gamma a_{1}^{\dag }a_{1}-i\gamma a_{N}^{\dag }a_{N},  \label{H}
\end{equation}%
where $a_{l}^{\dag }$ is the creation operator of the boson (or fermion) at $%
l$th site, the tunnelling strength and potential are denoted by $J$\ and $%
\pm i\gamma $. $\mathcal{P}$ and $\mathcal{T}$\ represent the
space-reflection operator, or parity operator and the time-reversal operator
respectively. The effects of $\mathcal{P}$ and $\mathcal{T}$\ on a discrete
system are

\begin{equation}
\mathcal{T}i\mathcal{T}=-i\text{, }\mathcal{P}a_{l}^{\dag }\mathcal{P}%
=a_{N+1-l}^{\dag }.
\end{equation}%
Obviously, the Hamiltonian Eq. (\ref{H}) has $\mathcal{PT}$ symmetry, i.e., $%
H^{\mathcal{PT}}=\mathcal{PT}H\mathcal{PT}=H$. According to the $\mathcal{PT}
$-symmetric\ quantum mechanics \cite{Bender 02}, $H$ can be further
classified to be either\ unbroken $\mathcal{PT}$ symmetry or broken $%
\mathcal{PT}$ symmetry, which depends on the symmetry of the eigenstates $%
\left\vert \psi _{k}\right\rangle $ in different regions of $\gamma $. The
time-independent Schr\"{o}dinger equation is%
\begin{equation}
H\left\vert \psi _{k}\right\rangle =\varepsilon _{k}\left\vert \psi
_{k}\right\rangle  \label{S eq1}
\end{equation}%
with corresponding eigenvalue $\varepsilon _{k}$. The system is unbroken $%
\mathcal{PT}$ symmetry if \textit{all} the eigenfunctions have $\mathcal{PT}$
symmetry%
\begin{equation}
\mathcal{PT}\left\vert \psi _{k}\right\rangle =\left\vert \psi
_{k}\right\rangle  \label{PT state}
\end{equation}%
and all the corresponding eigenvalues are real simultaneously. This
classification depends on the value of the parameter $\gamma $. Beyond the
unbroken $\mathcal{PT}$ symmetric region the system is broken $\mathcal{PT}$
symmetry, where Eq. (\ref{PT state}) does not hold for \textit{all} the
eigenfunctions and the eigenvalues of broken $\mathcal{PT}$ symmetric
eigenfunctions are imaginary. One of the aims of this paper is to provide
the complete exact eigenfunctions, eigenvalues,\ and the boundary between
unbroken and broken $\mathcal{PT}$ symmetric regions.

Acting the $\mathcal{PT}$\ operation on an arbitrary single-particle wave
function%
\begin{equation}
\left\vert \varphi \right\rangle =\sum_{l}h_{l}a_{l}^{\dag }\left\vert
0\right\rangle ,
\end{equation}%
where $\left\vert 0\right\rangle $\ denotes the vacuum state, we have%
\begin{equation}
\mathcal{PT}\left\vert \varphi \right\rangle =\sum_{l}\left( h_{l}\right)
^{\ast }a_{N+1-l}^{\dag }\left\vert 0\right\rangle =\sum_{l}\left(
h_{N+1-l}\right) ^{\ast }a_{l}^{\dag }\left\vert 0\right\rangle .
\end{equation}%
Then in the rest of this paper, we can simply use%
\begin{equation}
\mathcal{PT}h_{l}=\left( h_{N+1-l}\right) ^{\ast }
\end{equation}%
to present the $\mathcal{PT}$\ operation on the single-particle wavefunction.

\section{Bethe ansatz solutions}

In this paper, we only focus on the single-particle case. We denote the
single-particle eigenfunction in the form

\begin{equation}
\left\vert \psi _{k}\right\rangle =f_{k}^{l}a_{l}^{\dag }\left\vert
0\right\rangle .
\end{equation}
The Bethe ansatz wavefunction can be expressed in the form

\begin{equation}
f_{k}^{l}=A(k)e^{ikl}+B\left( k\right) e^{-ikl}.
\end{equation}%
where the quasi-momentum $k$\ and amplitudes $A(k)$, $B\left( k\right) $ can
be determined from the Eq. (\ref{S eq1}) and the proper definition of the
inner product according to the $\mathcal{PT}$-symmetric\ quantum theory \cite%
{Bender 02}. The solutions are presented explicitly in the following.

\subsection{Unbroken $\mathcal{PT}$-symmetric region}

In the unbroken $\mathcal{PT}$ symmetric region $\gamma \prec \gamma _{c}$
where
\begin{equation}
\gamma _{c}=\left\{
\begin{array}{c}
J\sqrt{\frac{n+1}{n}},\text{ }N=2n+1 \\
J,\text{ }N=2n%
\end{array}%
\right. ,  \label{gammaC}
\end{equation}%
the explicit $\mathcal{CPT}$ normalized wavefunctions are

\begin{equation}
f_{k}^{l}=\frac{e^{ik\left( l-N_{0}\right) }-\eta \left( k\right)
e^{-ik\left( l+N_{0}\right) }}{\left\vert \sqrt{\left[ 1+\left\vert \eta
\left( k\right) \right\vert ^{2}\right] \sin \left( Nk\right) /\sin k-2N\eta
\left( k\right) e^{-ik\left( N+1\right) }}\right\vert },  \label{real wf}
\end{equation}%
where $N_{0}=\left( N+1\right) /2$ is the center of the chain and the
coefficient%
\begin{equation}
\eta \left( k\right) =\frac{\gamma e^{ik}-iJ}{\gamma e^{-ik}-iJ}.
\label{aniso coefficient}
\end{equation}%
The quasi-momentum $k$\ satisfies the equation%
\begin{equation}
\gamma ^{2}\sin \left[ k(N-1)\right] +J^{2}\sin \left[ k(N+1)\right] =0
\label{real eq}
\end{equation}%
which has $N$ real solutions. All the corresponding eigenvalues are real%
\begin{equation}
\varepsilon _{k}=-2J\cos k,  \label{real eigenvalue}
\end{equation}%
with the quasi-momentum $k$ being more explicit form%
\begin{equation}
k=\frac{n_{k}\pi +\theta _{k}}{N},n_{k}\in \lbrack 1,N]  \label{k}
\end{equation}%
and%
\begin{equation}
\theta _{k}=\tan ^{-1}\left[ \frac{\left( \gamma ^{2}-J^{2}\right) }{\left(
\gamma ^{2}+J^{2}\right) }\tan k\right] .  \label{theta_k}
\end{equation}%
The reality of the spectrum is a consequence of $\mathcal{PT}$ invariance.
At $\gamma =0$,\ we have $\eta \left( k\right) =1$,\ the eigenfunctions
reduce to the form

\begin{eqnarray}
f_{l}^{k}(\gamma &=&0)=\left( -i\right) ^{n_{k}}i\sqrt{\frac{2}{N+1}}\sin
\left( kl\right) ,  \label{standing wave} \\
k &=&\frac{\pi n_{k}}{N+1},n_{k}\in \lbrack 1,N],  \notag
\end{eqnarray}%
which is the well-known solution of a Hermitian tight-binding chain.

\subsection{Broken $\mathcal{PT}$-symmetric region}

In the region $\gamma \succ \gamma _{c}$, the $\mathcal{PT}$ symmetry of the
Hamiltonian is spontaneously broken; even though $\mathcal{PT}$ commutes
with $H$, the eigenfunctions of $H$ are not \textit{all} simultaneously
eigenfunctions of $\mathcal{PT}$. In this region, it can be shown that there
are $N-2$ real $k$ for the equation Eq. (\ref{real eq}) which corresponds to
the eigenfunctions Eq. (\ref{real wf})\ and real eigenvalues Eq. (\ref{real
eigenvalue}). The rest two eigenfunctions correspond complex quasi-momenta,
which are in the form $k=\pi /2\pm i\left\vert \kappa \right\vert $ with $%
\kappa $\ satisfying the equation%
\begin{equation}
\begin{array}{c}
\gamma ^{2}\sinh \left[ \kappa (N-1)\right] =J^{2}\sinh \left[ \kappa (N+1)%
\right] \text{, }\left( \text{odd }N\right) ; \\
\gamma ^{2}\cosh \left[ \kappa (N-1)\right] =J^{2}\cosh \left[ \kappa (N+1)%
\right] \text{, }\left( \text{even }N\right) .%
\end{array}
\label{imaginary eq}
\end{equation}%
The broken $\mathcal{PT}$-symmetric eigen functions can be written as%
\begin{equation}
f_{\pi /2\pm i\kappa }^{l}\varpropto e^{\pm \kappa N_{0}}\left[ \left(
i\right) ^{l}e^{\mp \kappa l}-\left( -i\right) ^{l}\frac{J-\gamma e^{\mp
\kappa }}{J+\gamma e^{\pm \kappa }}e^{\pm \kappa l}\right]
\label{imaginary wfs}
\end{equation}%
with the imaginary eigenvalues%
\begin{equation}
\varepsilon _{k}=\pm i2J\sinh \kappa .  \label{imaginary eigenvalue}
\end{equation}%
At this stage, the wave functions Eq. (\ref{imaginary wfs}) are not
normalized since the standard $\mathcal{CPT}$ normalization procedure is
invalid for such broken $\mathcal{PT}$-symmetric wave functions.

\subsection{$\mathcal{CPT}$ formalism}

In this subsection, we will elucidate the $\mathcal{CPT}$ formalism based on
the solutions of the present model. It can be seen that the solutions have
different features compared to that of a Hermitian Hamiltonian. The most
intuitive novelty of the eigenfunctions Eq. (\ref{real wf}) with real
eigenvalues are not standing waves due to $\left\vert \eta \left( k\right)
\right\vert $ being not unitary for nonzero $\gamma $. According to the $%
\mathcal{PT}$-symmetric\ quantum theory \cite{Bender 02}, the correct inner
product is determined by the Hamiltonian itself, so\ it is necessary for the
discrete system Eq. (\ref{H}) to establish a self-consistent formalism based
on the obtained exact solutions.

A straightforward calculation shows that the eigenfunctions Eq. (\ref{real
wf}) obey the $\mathcal{PT}$-symmetry%
\begin{equation}
\mathcal{PT}f_{k}^{l}=f_{k}^{l}.
\end{equation}%
According to the $\mathcal{PT}$-symmetry quantum theory \cite{Bender 02},
one can define the coordinate-space representations of the parity operator $%
\mathcal{P}$\ and $\mathcal{C}$ operator for a discrete system as%
\begin{eqnarray}
\mathcal{P}(m,l) &=&\delta _{m,N+1-l}, \\
\mathcal{C}(m,l) &=&\sum_{k}f_{k}^{m}f_{k}^{l},
\end{eqnarray}%
which lead to the $\mathcal{CPT}$\ orthogonal and normalization relations

\begin{equation}
\sum_{l,m}f_{k}^{l}\mathcal{C}(l,m)\mathcal{P}(m,l)\left( f_{k^{\prime
}}^{l}\right) ^{\ast }=\delta _{kk^{\prime }}.  \label{orth renorm}
\end{equation}%
This guarantees the positive-definite $\mathcal{CPT}$\ inner product of two
arbitrary states and that the time evolution is unitary. One can verify that
the operator $\mathcal{C}$ satisfies

\begin{equation}
\mathcal{C}^{2}=1\text{, }\left[ \mathcal{C},\mathcal{PT}\right] =0\text{, }%
\left[ \mathcal{C},H\right] =0.
\end{equation}%
In the next section, the above formalism will be utilized to establish the
canonical basis to construct the equivalent Hermitian Hamiltonian in the
unbroken $\mathcal{PT}$-symmetric region. We will see that choosing $%
\mathcal{CPT}$\ normalized eigenstates of H leads the Hermitian equivalence
matrix\ to be more symmetrical.

\section{Exceptional points}

In this section, we investigate the critical behavior of the system as $%
\gamma $ in the vicinity of $\gamma _{c}$. From the obtained solutions of
the model, the critical point $\gamma _{c}$ has the characteristics of
exceptional point \cite{Heiss,Heiss94,Heiss37,HeissPRL}.\ We will
investigate the feature of eigenvalues and eigenfunctions around the
exceptional point in detail.

In unbroken $\mathcal{PT}$-symmetric region, as $\gamma $ approaches to $%
\gamma _{c}$,\ the solutions of Eq. (\ref{real eq}) change abruptly.
Actually, as the function%
\begin{equation}
\mathcal{F}(k)=\gamma ^{2}\sin \left[ k(N-1)\right] -J^{2}\sin \left[ k(N+1)%
\right]
\end{equation}%
is an even (odd) function about $k=\pi /2$ for even (odd)$\ N$, two real
energy levels, which are closest to zero, disappear when $\gamma $ passes
through the boundary $\gamma _{c}$ of the two regions. Meanwhile, two
imaginary energy\ levels appear. Thus the critical behavior of the
eigenstates can be characterized by the two components problem. For $\gamma
=\gamma _{c}-0^{+}$, we denote the referring eigenfunctions as $f_{\pi
/2+\delta }^{l}$ and $f_{\pi /2-\delta }^{l}$ with
\begin{eqnarray}
\delta  &\simeq &\left\{
\begin{array}{cc}
\frac{1}{\sqrt{-N\alpha }}, & N\text{ is even} \\
\sqrt{\frac{3\left( \alpha -N\right) }{N^{3}-\alpha }}, & N\text{ is odd}%
\end{array}%
\right. ,  \label{delta} \\
\alpha  &=&\frac{J^{2}+\gamma ^{2}}{\gamma ^{2}-J^{2}},
\end{eqnarray}%
It follows that such two eigenfunctions approach to a same function and
their $\mathcal{PT}$ norms\ tend to zero when $\mathcal{PT}$ symmetry is
broken.

In the broken $\mathcal{PT}$-symmetric region, for $\gamma =\gamma
_{c}+0^{+} $,\ such two eigenfunctions are replaced by Eq. (\ref{imaginary
wfs}), while the corresponding eigenvalues turn to imaginary values Eq. (\ref%
{imaginary eigenvalue}). Other$\ N-2$ eigenfunctions in the form of Eq. (\ref%
{real wf}) with real eigenvalues still satisfy the $\mathcal{CPT}$
orthogonal and normalized relations Eq. (\ref{orth renorm}). However, the
two eigenfunctions Eq. (\ref{imaginary wfs}) can no longer be normalized via
the above $\mathcal{CPT}$ inner product since they have the following
features%
\begin{equation}
\mathcal{PT}\left[ f_{\pi /2+i\kappa }^{l}\right] \varpropto f_{\pi
/2-i\kappa }^{l}
\end{equation}%
and%
\begin{equation}
f_{\pi /2\pm i\kappa }^{l}\mathcal{PT}\left[ f_{\pi /2\pm i\kappa }^{l}%
\right] =0.
\end{equation}

\begin{figure}[tp]
\includegraphics[ bb=16 181 554 604, width=6 cm, clip]{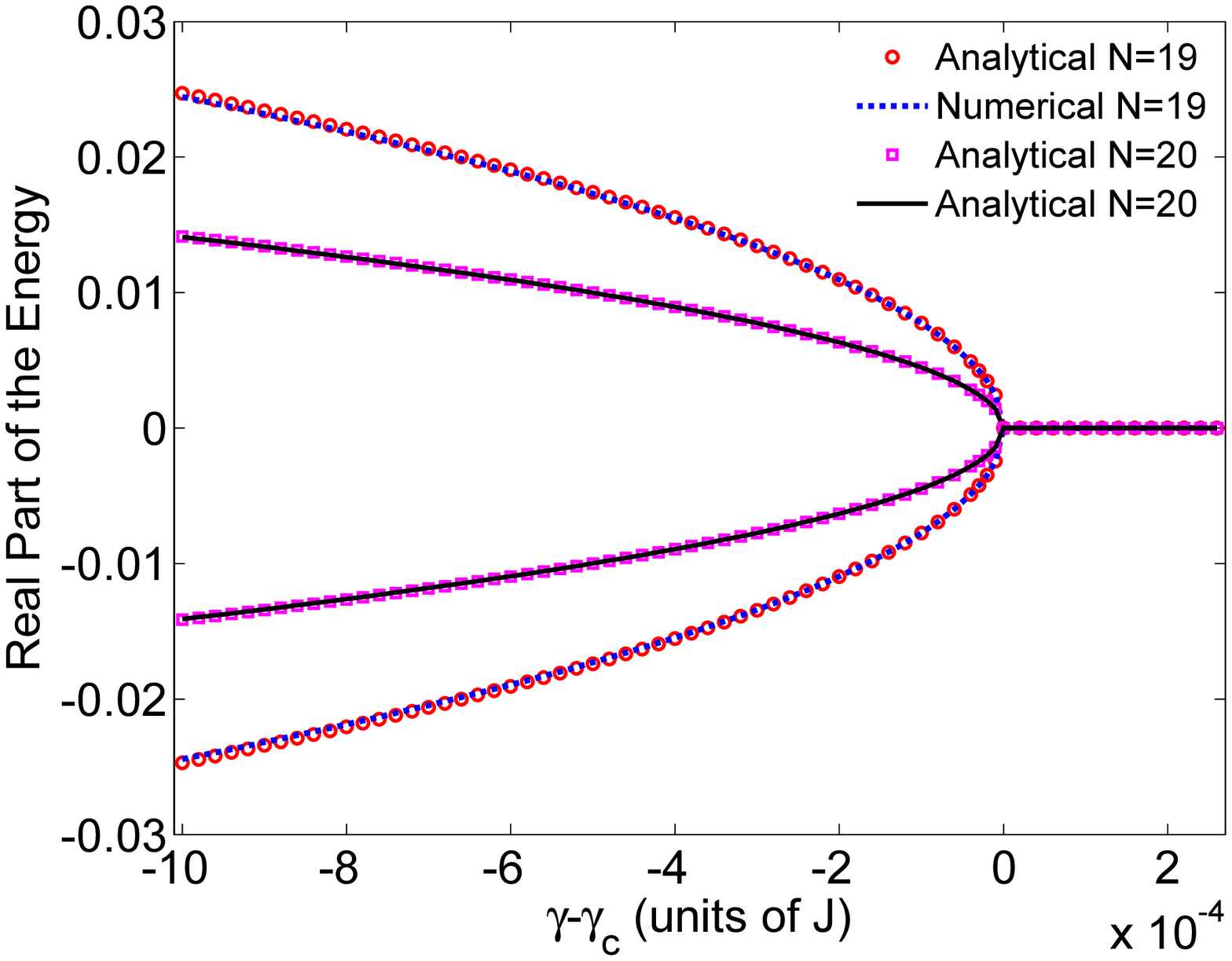} %
\includegraphics[ bb=16 181 554 604, width=6 cm, clip]{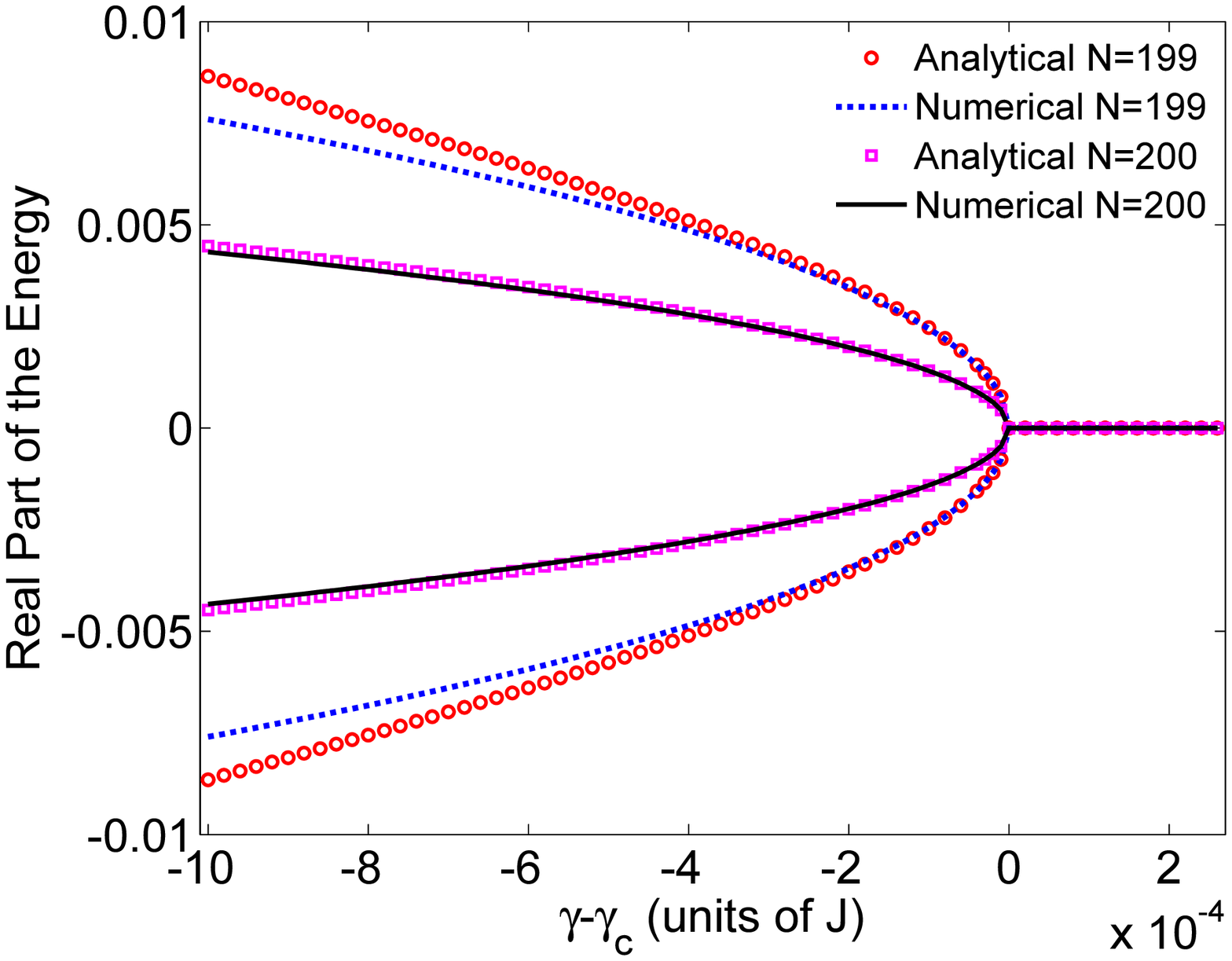} %
\includegraphics[ bb=16 190 554 604, width=6 cm, clip]{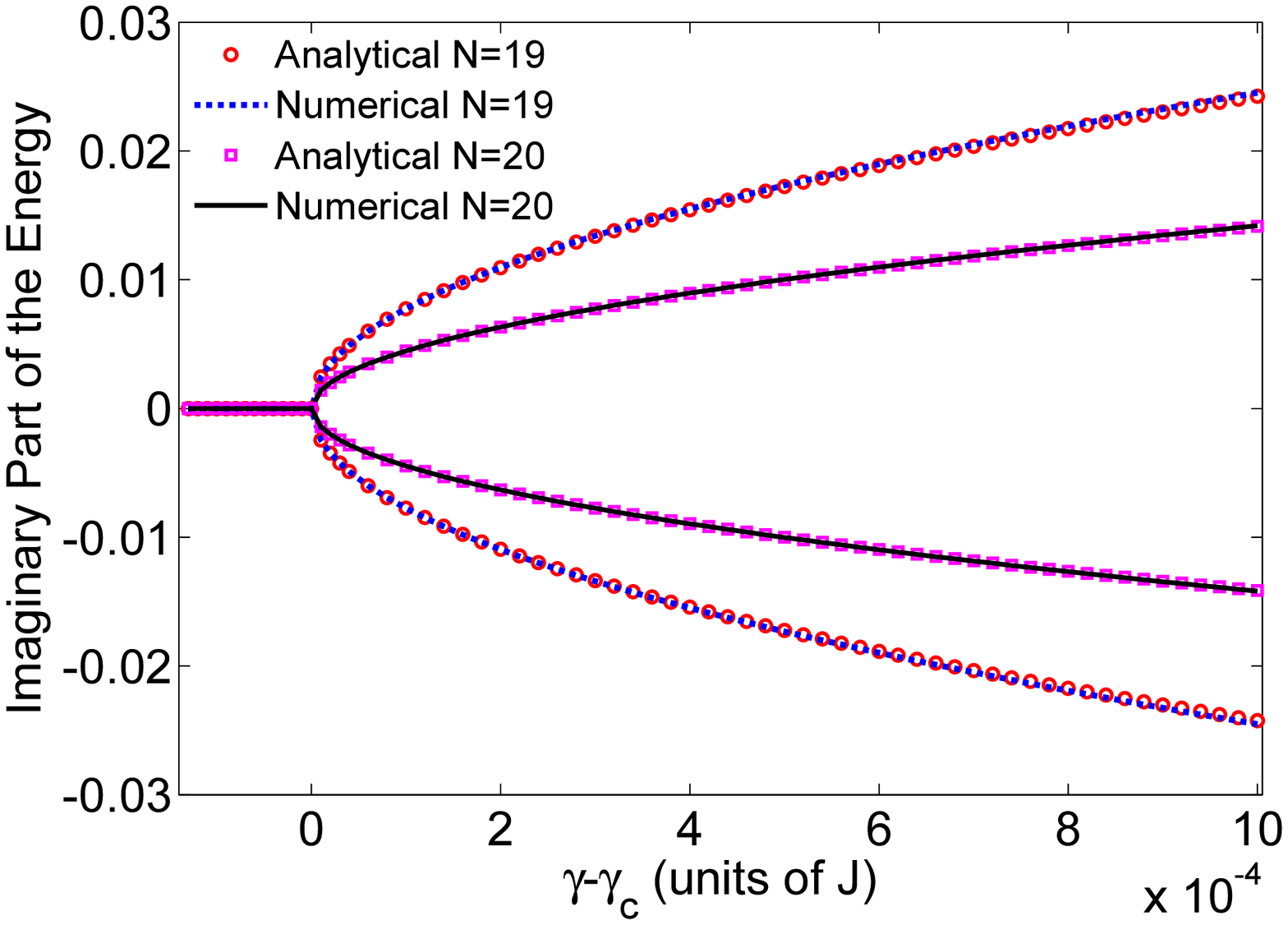} %
\includegraphics[ bb=16 181 554 604, width=6 cm, clip]{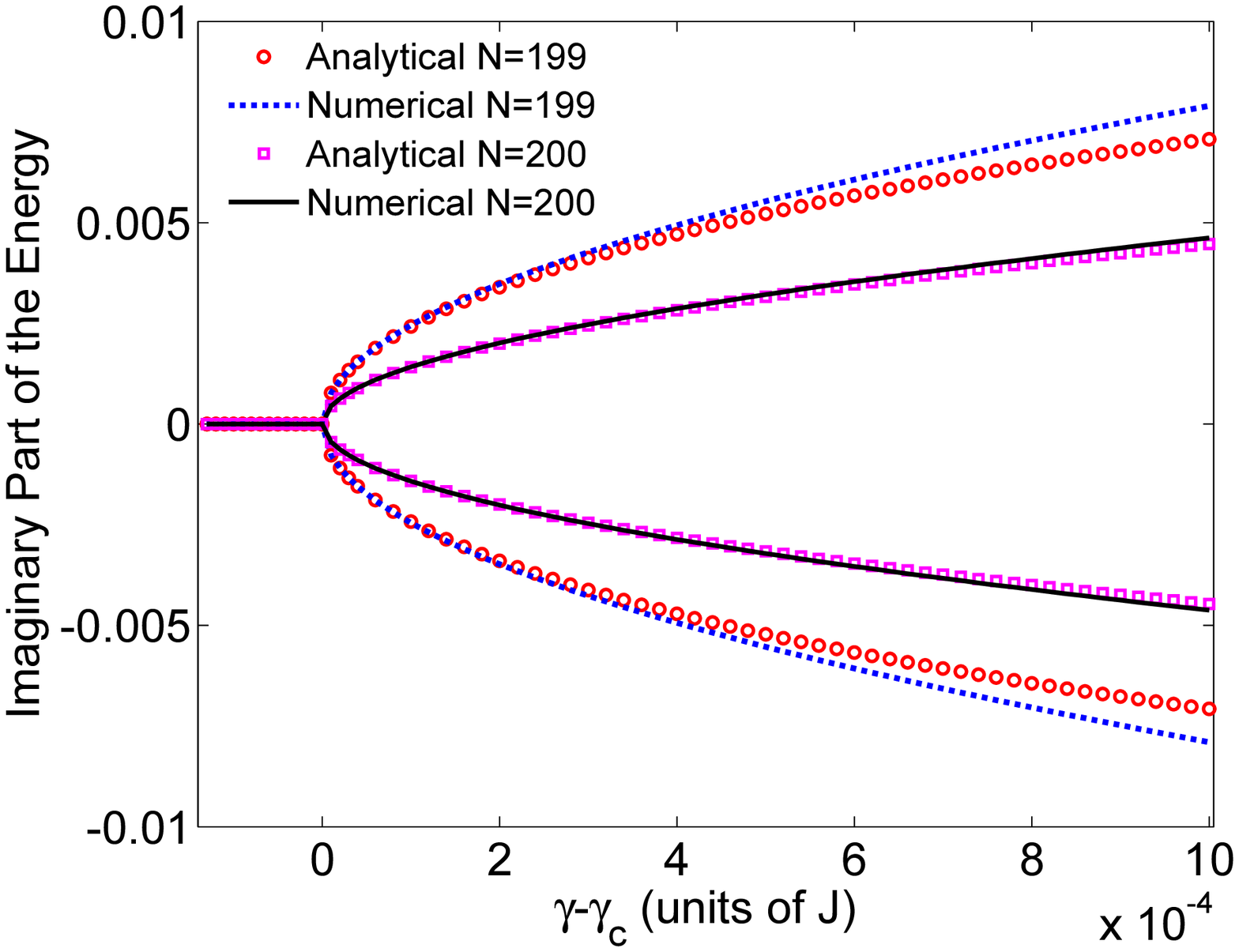}
\caption{(Color online) Real and imaginary parts of two repelling levels as
functions of $\protect\gamma -\protect\gamma _{c}$. The plots are obtained
from the approximate analytical results Eqs. (\protect\ref{E_left}), (%
\protect\ref{E_right}) and numerical simulations for the systems with $N=19$%
, $20$, $199$, and $200$. It shows that the analytical eigenvalue
expressions are good approximation to the numerically exact results,
especially for small $N$ system.}
\label{fig1}
\end{figure}

On the other hand, the corresponding energy levels experience a\ switch from
real to complex values as well as a\ coalescence at the exceptional point.
In the unbroken symmetric side, from (\ref{delta}) we have%
\begin{eqnarray}
\varepsilon _{\pi /2\pm \delta } &\simeq &\pm 2J\sin \delta  \label{E_left}
\\
&\simeq &\pm 2J\delta .  \notag
\end{eqnarray}%
In the broken symmetric side, from (\ref{imaginary eq}) we have

\begin{equation}
\kappa \simeq \left\{
\begin{array}{cc}
\frac{1}{\sqrt{N\alpha }}, & N\text{ is even} \\
\sqrt{\frac{3\left( N-\alpha \right) }{N^{3}-\alpha }}, & N\text{ is odd}%
\end{array}%
\right. ,
\end{equation}%
which correspond to the eigenvalues%
\begin{eqnarray}
\varepsilon _{\pi /2\pm i\kappa } &\simeq &\pm 2iJ\sinh \kappa
\label{E_right} \\
&\simeq &\pm 2iJ\kappa .  \notag
\end{eqnarray}%
Alternatively, taking $\left\vert \gamma -\gamma _{c}\right\vert $ as the
variable, we can see that the two concerned eigen states satisfy%
\begin{equation}
\varepsilon _{\pi /2\pm \delta }\simeq \text{Im}(\varepsilon _{\pi /2\pm
i\kappa })\simeq \left\{
\begin{array}{cc}
\pm 2J\sqrt{\frac{\left\vert \gamma -\gamma _{c}\right\vert }{N\gamma _{c}}},
& N\text{ is even} \\
\pm 2J\sqrt{\frac{3\left\vert \gamma -\gamma _{c}\right\vert }{N\gamma _{c}}}%
, & N\text{ is odd}%
\end{array}%
\right. ,
\end{equation}%
near the critical point, which reveals the symmetry of the critical
behavior. In Fig. (\ref{fig1}) we plot the real and imaginary parts of two
repelling levels as functions of $\gamma -\gamma _{c}$, which are obtained
from the approximate analytical results Eqs. (\ref{E_left}), (\ref{E_right})
and numerical simulations for finite systems. The analytical eigenvalue
expressions (\ref{E_left}) and (\ref{E_right})\ are good approximations to
the numerically exact results.\ It shows that the exceptional points are
always associated with a level repulsion in the vicinity of them. The
square-root type functions for the energy reveal this characteristics.
Previous study \cite{Heiss94,Heiss37}, shows that the exceptional point is
often related to the emergence of chaotic behavior.\ However, the quantum
chaos is not found in the present model. This may be due to that the
coalescence and repulsion of levels in this model only relate to two
eigenstates rather than multi-level. Thus there is no occurrence of quantum
chaos.

%
%%%%%%%%%%%%%%%%%%%%%%%%%%%%%%%%%%%%%%%%%%%%%%%%%%%%%%%%%%%%%%%%%%%%%%%%
%%%%%%%%%%%%%%%%%%%%%%%%%%%%%%%%%%%%%%%%%%%%%%%%%%%%%%%%%%%%%%%%%%%%%%%%%

\section{Equivalent Hermitian Hamiltonian}

A natural question to ask is whether such a model Eq. (\ref{H}) describes
real physics or is just an unrealistic mathematical product. While there is
as yet no answer to this question, we can gain some insight regarding its
equivalent Hermitian counterpart \cite{A.M36,Jones}. This section aims at
seeking the equivalent Hermitian Hamiltonian\ of the present model in the
framework of metric-operator theory \cite{A.M}.

We start with the eigenstates of the Hamiltonian $H^{\dag }=H^{\ast
}=H(-\gamma )$. From (\ref{real wf}) the explicit $\mathcal{CPT}$ normalized
wavefunctions of $H^{\dag }$\ within the unbroken $\mathcal{PT}$-symmetric
region\ are%
\begin{equation}
g_{k}^{l}=\frac{e^{ik\left( l-N_{0}\right) }-\zeta \left( k\right)
e^{-ik\left( l+N_{0}\right) }}{\left\vert \sqrt{\left[ 1+\left\vert \zeta
\left( k\right) \right\vert ^{2}\right] \sin \left( Nk\right) /\sin
k-2N\zeta \left( k\right) e^{-ik\left( N+1\right) }}\right\vert },
\label{g wf}
\end{equation}%
where the coefficient is

\begin{equation}
\zeta \left( k\right) =\frac{\gamma e^{ik}+iJ}{\gamma e^{-ik}+iJ},
\end{equation}%
and the corresponding eigenvalues are (\ref{real eigenvalue}), the same as
that of $f_{k}^{l}$. Within the unbroken $\mathcal{PT}$-symmetric region, we
have

\begin{equation}
\mathcal{PT}\left\vert g_{k}\right\rangle =\left\vert g_{k}\right\rangle ,
\label{g_PT}
\end{equation}%
and two sets of eigenfunctions $\left\{ g_{k}^{l}\right\} $\ and $\left\{
f_{k}^{l}\right\} $\ form a biorthonormal system, i.e.,%
\begin{equation}
\sum_{l}\left( g_{k}^{l}\right) ^{\ast }f_{k^{\prime }}^{l}=\delta
_{kk^{\prime }}.  \label{biorthonormal system}
\end{equation}

According to the metric-operator theory \cite{A.M}, one can construct a
positive-definite operator

\begin{equation}
\eta _{+}=\sum_{k}\left\vert g_{k}\right\rangle \left\langle
g_{k}\right\vert .  \label{metric operator}
\end{equation}%
From the Appendix, it can be shown that this operator satisfies%
\begin{eqnarray}
\left( \eta _{+}\right) ^{\dag } &=&\eta _{+},  \label{eta_hermiticity} \\
\left( \eta _{+}\right) ^{-1} &=&\left( \eta _{+}\right) ^{\ast },
\label{eta_conjugate} \\
\mathcal{PT}\eta _{+}\mathcal{PT} &=&\eta _{+}.  \label{eta_PT}
\end{eqnarray}%
In the spatial coordinate space spanned by\ basis $\left\vert m\right\rangle
=a_{m}^{\dag }\left\vert 0\right\rangle $, the matrix representation of $%
\eta _{+}$ has the form%
\begin{equation}
\left\langle m\right\vert \eta _{+}\left\vert n\right\rangle =\sum_{k}\left(
g_{k}^{m}\right) ^{\ast }g_{k}^{n}.
\end{equation}%
More explicitly, it can be shown that the matrix has the following properties%
\begin{equation}
\left\langle m\right\vert \eta _{+}\left\vert n\right\rangle =\left\langle
N+1-n\right\vert \eta _{+}\left\vert N+1-m\right\rangle .
\end{equation}%
and%
\begin{equation}
\left\langle m\right\vert \eta _{+}\left\vert n\right\rangle
=(-1)^{m+n}\left\langle m\right\vert \eta _{+}\left\vert n\right\rangle
^{\ast }.  \label{eta_elements}
\end{equation}%
In addition, defining the canonical transformation $R$,

\begin{equation}
R\left\vert l\right\rangle =(-1)^{l}\left\vert l\right\rangle ,
\label{R transformation}
\end{equation}%
we have

\begin{equation}
R\eta _{+}R^{-1}=\left( \eta _{+}\right) ^{\ast }=\left( \eta _{+}\right)
^{-1}.  \label{R_conjugate}
\end{equation}%
These features allow characterizing the equivalent physical system of $H$.\
In the eigenspace of $\eta _{+}$, we can rewrite $\eta _{+}$\ as

\begin{equation}
\eta _{+}=\sum_{n}\epsilon _{n}\left\vert \epsilon _{n}\right\rangle
\left\langle \epsilon _{n}\right\vert
\end{equation}%
where $\left\vert \epsilon _{n}\right\rangle $ is the eigenvector of
operator $\eta _{+}$ with the eigenvalue $\epsilon _{n}$, i.e.,

\begin{equation}
\eta _{+}\left\vert \epsilon _{n}\right\rangle =\epsilon _{n}\left\vert
\epsilon _{n}\right\rangle .
\end{equation}%
The eigenvalues $\epsilon _{n}$ are all real due to the Hermiticity of $\eta
_{+}$ and the complete set $\left\{ \left\vert \epsilon _{n}\right\rangle
\right\} $ is referred as a canonical metric basis.

Accordingly, the equivalent Hermitian Hamiltonian $\mathcal{H}$ can be
obtained by a unitary transformation and can be expressed as%
\begin{equation}
\mathcal{H}=\sum\limits_{m,n}\sqrt{\frac{\epsilon _{m}}{\epsilon _{n}}}%
H_{mn}\left\vert \epsilon _{m}\right\rangle \left\langle \epsilon
_{n}\right\vert  \label{h}
\end{equation}%
where $H_{mn}=\left\langle \epsilon _{m}\right\vert H\left\vert \epsilon
_{n}\right\rangle $ is the matrix representation of Hamiltonian $H$ under
this canonical metric basis $\left\{ \left\vert \epsilon _{n}\right\rangle
\right\} $. The Hermitian equivalence matrix $\mathcal{H}$ can be achieved as

\begin{equation}
\mathcal{H}_{mn}=\sqrt{\frac{\epsilon _{m}}{\epsilon _{n}}}%
\sum\limits_{i,j}\left\langle i\right\vert H\left\vert j\right\rangle \left(
\epsilon _{m}^{i}\right) ^{\ast }\epsilon _{n}^{j},
\end{equation}%
where $\epsilon _{m}^{i}=\left\langle i\right. \left\vert \epsilon
_{m}\right\rangle $ is the component of state $\left\vert \epsilon
_{m}\right\rangle $\ in the spatial coordinate basis $\left\vert
i\right\rangle $.

\begin{figure}[tp]
\includegraphics[ bb=42 292 456 750, width=4 cm, clip]{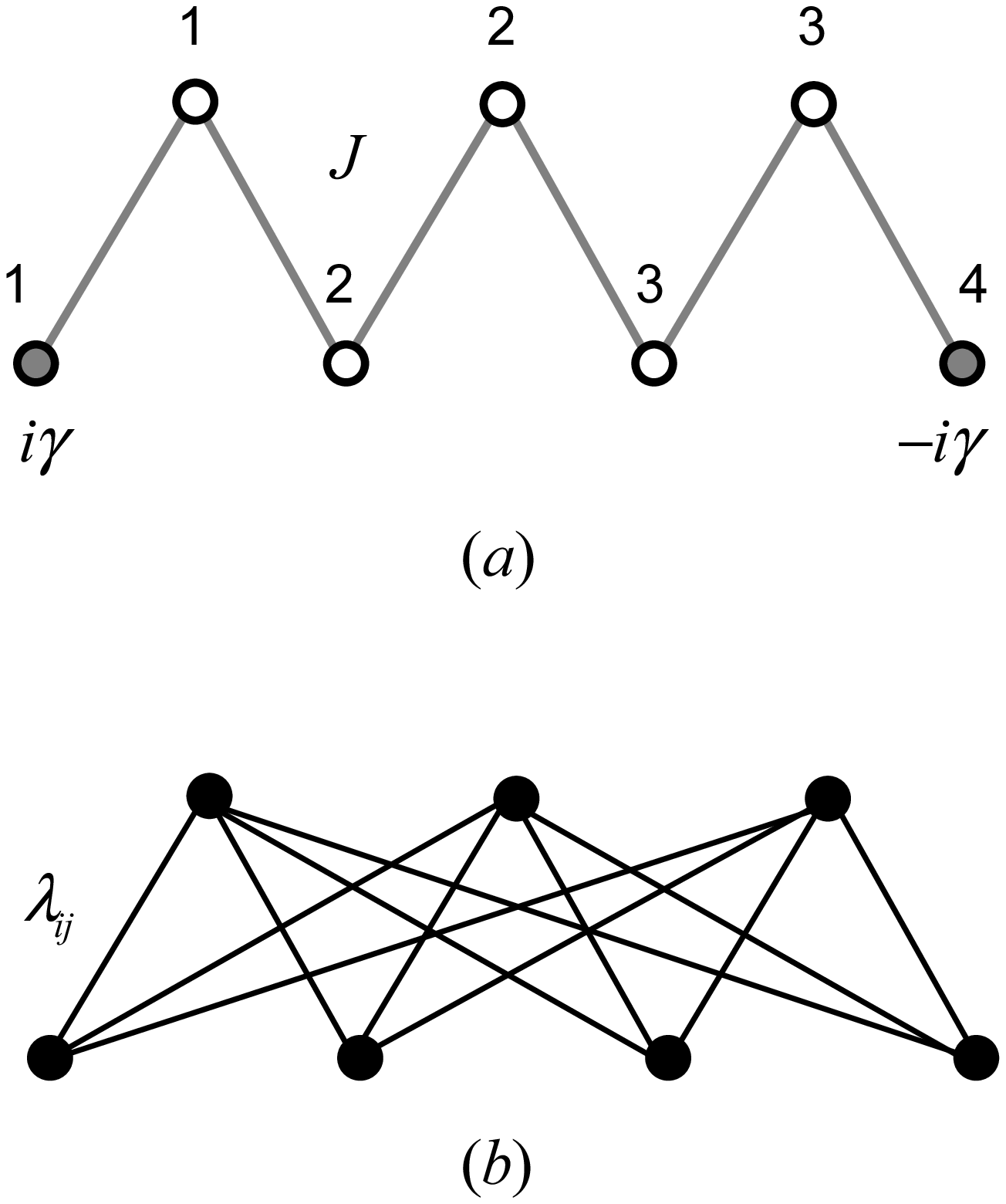} %
\includegraphics[ bb=42 292 456 750, width=4 cm, clip]{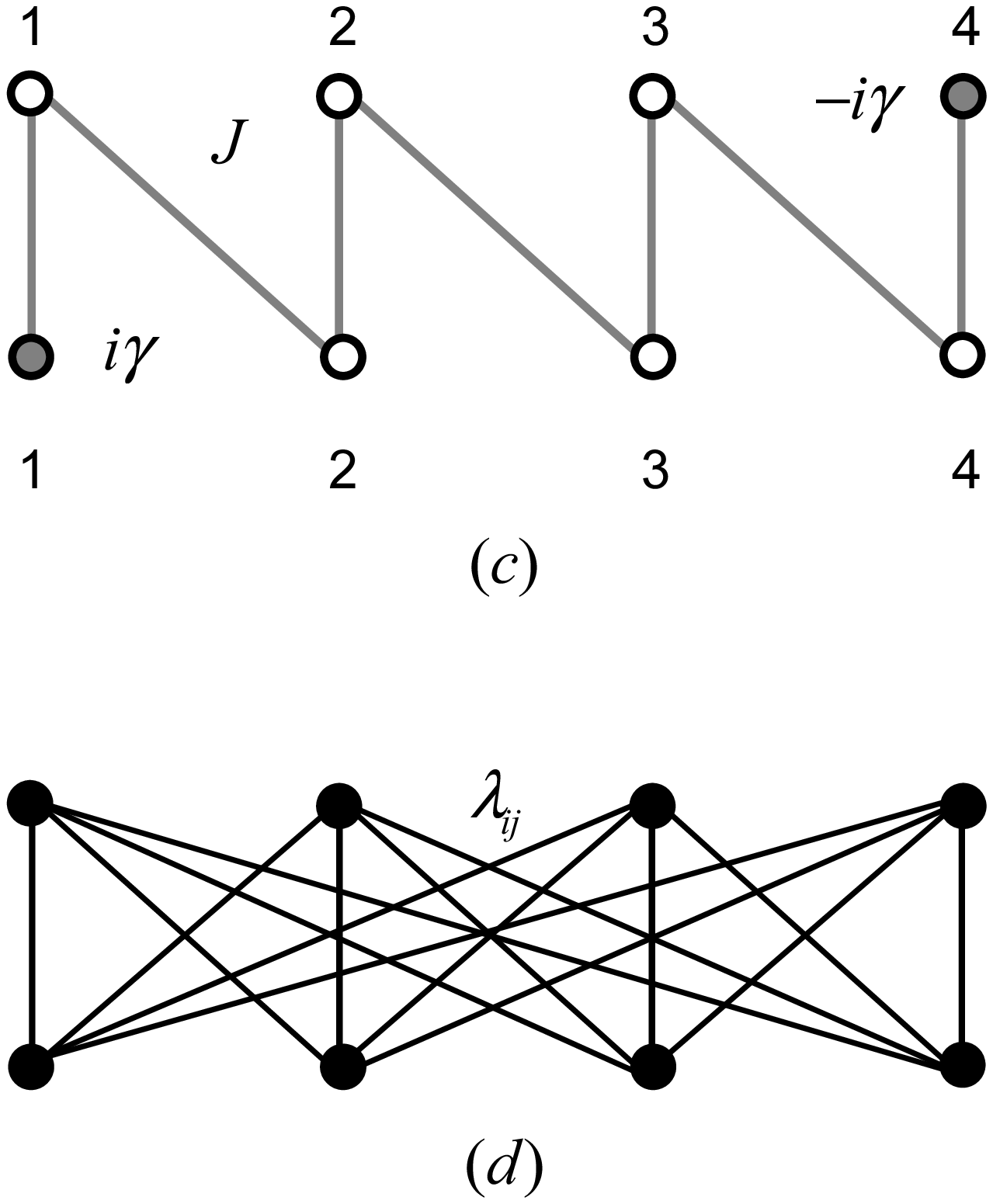}
\caption{{}(Color online) Schematic illustration of configurations for (a,c)
pseudo-hermitian system $H$ on $7$, $8$-site lattices and (b,d) their
hermitian counterparts $h$.\ Both the original system and its hermitian
counterpart are all bipartite graphs.}
\label{graphs}
\end{figure}

Before we explore the features of the equivalent Hermitian Hamiltonian $%
\mathcal{H}$, it is worthy to point that the ordinal tight-binding chain (%
\ref{H}) is a bipartite lattice which can be separated into $A$ and $B$
sublattices with sites $N_{A}$\ and $N_{B}$, respectively. We have $%
N=N_{A}+N_{B}$ with $\left\vert N_{A}-N_{B}\right\vert =1$ ($0$)\ for odd
(even) $N$ and the two sublattices are connected by hopping terms with
strength $J$ (see Fig. \ref{graphs} (a) and (c)). Now we turn to the
properties of $\mathcal{H}$. It can be shown, from the Appendix of this
paper \cite{remark}, that under a proper choice of the transformation of $%
\left\{ \left\vert \epsilon _{n}\right\rangle \right\} $, the matrix
representation of $\mathcal{H}$ can be in the form of

\begin{equation}
\mathcal{H}=\left[
\begin{array}{cc}
0 & A \\
A^{T} & 0%
\end{array}%
\right]  \label{h_matrix}
\end{equation}%
where $A^{T}$ is the transposed matrix of a $N_{A}\times N_{B}$ matrix $A$,
and $A$ satisfies
\begin{equation}
A_{ij}=\left\{
\begin{array}{cc}
A_{N_{A}+1-j,N_{B}+1-i}, & N\text{ is even} \\
A_{N_{A}+1-i,N_{B}+1-j}, & N\text{ is odd}%
\end{array}%
\right. .  \label{antisymmetry}
\end{equation}%
Then real symmetric matrix $\mathcal{H}$\ can be regarded as a
single-particle matrix representation of a tight-binding model on a
bipartite lattice $N=N_{A}+N_{B}$ with $\left\vert N_{A}-N_{B}\right\vert =1$
($0$)\ for odd (even) $N$.\ Two sublattices are connected by\ the long-range
hopping terms with strength $\lambda _{ij}$. The corresponding tight-binding
Hamiltonian can be expressed as%
\begin{equation}
\mathcal{H}=\sum_{i=1}^{N_{A}}\sum_{j=1}^{N_{B}}\lambda _{ij}\left(
\left\vert i\right\rangle _{AB}\left\langle j\right\vert +\text{H.c.}\right)
,  \label{h_tb}
\end{equation}%
where $\left\vert l\right\rangle _{A}$ and $\left\vert l\right\rangle _{B}\ $%
denote single-particle states referring to $l$th site in sublattices $N_{A}$%
\ and $N_{B}$, respectively. Note that both the original model $H$ and its
Hermitian counterpart $\mathcal{H}$\ are all bipartite. The former contains
NN couplings, while the latter contains the long-range couplings. This fact
agrees with the observations from other example Hamiltonians, that in
general the Hermitian counterpart of a pseudo-Hermitian Hamiltonian obtained
by the metric-operator theory is nonlocal operator \cite{A.M}.

In order to exemplify the above analysis, we consider the small systems with
$N=7$ and $8$. Fig. \ref{graphs} shows the schematics of configurations for
(a,c) pseudo-Hermitian system $H$ on $7$, $8$-site lattices and (b,d) their
Hermitian counterparts $\mathcal{H}$. The corresponding hopping constants $%
\lambda _{ij}$ are computed for $\gamma =0.00,$ $0.50$, and $0.99$, which
are listed in Table 1 (a) and (b). It indicates that all the constants vary
within a narrow range without changing signs as $\gamma $\ covers the whole
unbroken symmetric region and obey the relation of Equation (\ref%
{antisymmetry}).

\begin{widetext}

\begin{center}
Table 1 (a)

\begin{tabular}{ccccccc}
\hline\hline
$\gamma $ &  $\ \lambda _{11},\lambda _{34}$  &  $\lambda _{12},\lambda _{33}
$  &  $\lambda _{13},\lambda _{32}$  &  $\lambda _{14},\lambda _{31}$  &  $%
\lambda _{21},\lambda _{24}$  &  $\lambda _{22},\lambda _{23}$  \\ \hline
$0.00$ & $0.6242$ & $1.0068$ & $-0.2997$ & $0.0830$ & $0.2071$ & $-1.2071$
\\
$0.50$ & $0.5703$ & $0.9731$ & $-0.3089$ & $0.0883$ & $0.2039$ & $-1.2075$
\\
$0.99$ & $0.3355$ & $0.8949$ & $-0.3280$ & $0.0774$ & $0.1468$ & $-1.2089$
\\ \hline\hline
\end{tabular}

Table 1 (b)

\begin{tabular}{ccccccccccc}
\hline\hline
$\gamma $ &  $\ \lambda _{11},\lambda _{44}$  &  $\lambda _{12},\lambda _{34}
$  &  $\lambda _{13},\lambda _{24}$  & $\lambda _{14}$ &  $\lambda
_{21},\lambda _{43}$  &  $\lambda _{22},\lambda _{33}$  & $\lambda _{23}$ &
$\lambda _{31},\lambda _{42}$  & $\lambda _{32}$ & $\lambda _{41}$ \\ \hline
$0.00$ & $0.5627$ & $-0.9300$ & $-0.2994$ &  $-0.1199$  & $0.1954$ & $1.1615$
&  $-1.2411$  & $0.0914$ &  $0.3333$ \  & $0.0277$ \\
$0.50$ & $0.5153$ & $-0.8918$ & $-0.3057$ & $-0.1304$ & $0.1909$ & $1.1527$
& $-1.2469$ & $0.0972$ & $0.3461$ & $0.0310$ \\
$0.99$ & $0.1766$ & $-0.9005$ & $-0.3157$ & $-0.1458$ & $0.1005$ & $1.0333$
& $-1.3522$ & $0.0627$ & $0.3505$ & $0.0143$ \\ \hline\hline
\end{tabular}
\end{center}

\textit{Table 1. The coupling\ constant distributions }$\lambda _{ij}$
\textit{for systems with (a) }$N=7$\textit{\ and (b) }$8$\textit{,} \textit{%
obtained from numerical simulation for }$\gamma =0.00,$ $0.50$, and $0.99$.

\end{widetext}

%%%%%%%%%%%%%%%%%%%%%%%%%%%%%%%%%%%%%%%%%%%%%%%%%%%%%%%%%%%%%%%%%%%%%%%%
%%%%%%%%%%%%%%%%%%%%%%%%%%%%%%%%%%%%%%%%%%%%%%%%%%%%%%%%%%%%%%%%%%%%%%%%%

\section{Conclusion and discussion}

In conclusion, we have studied the Non-Hermitian quantum mechanics for the
discrete system. The exact analytic single-particle solution for a
tight-binding chain with two end imaginary potentials is obtained, which
substantiates the formalism of the $\mathcal{PT}$ quantum theory for
discrete system: Based on the Bethe ansatz results, it is found that, in
single-particle subspace, this model exhibits two phases, an unbroken
symmetry phase with a purely real energy spectrum in the region and a
spontaneously-broken symmetry phase with a pair of imaginary eigenvalues in
the region.

In addition, we have indicated boundary of two phases possesses the
characteristics of exceptional point. We also construct the equivalent
Hermitian Hamiltonian\ of the present model in the framework of
metric-operator theory. We find out that the equivalent Hermitian
Hamiltonian can be written as another bipartite lattice model with real
long-range hoppings.

It is worthwhile to note that the Non-Hermitian Hamiltonian (\ref{H})\
cannot simply be written as the form $\sum_{k}\epsilon _{k}a_{k}^{\dag
}a_{k} $, where $a_{k}^{\dag }=\sum_{l}f_{k}^{l}a_{l}^{\dag }$ is creation
operator in $k$ space, since the transition matrix is no longer unitary.
However, the equivalent Hermitian Hamiltonian $\mathcal{H}$ can be in such
form. It might be interesting to explore the multiparticle sector of the
system using the equivalent Hermitian Hamiltonian.

We acknowledge the support of the CNSF (Grants No. 10874091 and No.
2006CB921205).

\section{Appendix: Properties of matrices $\protect\eta _{+}$\ and $\mathcal{%
H}$}

According to the defination of the metric operator (\ref{metric operator})
and the properties of eigenfunctions, equations (\ref{g_PT}) and (\ref%
{biorthonormal system}), we have

\begin{eqnarray}
\left( \eta _{+}\right) ^{\ast }\eta _{+} &=&\left( \sum_{k}\left\vert
g_{k}\right\rangle \left\langle g_{k}\right\vert \right) ^{\ast
}\sum_{k^{\prime }}\left\vert g_{k^{\prime }}\right\rangle \left\langle
g_{k^{\prime }}\right\vert \\
&=&\sum_{k}\left\vert f_{k}\right\rangle \left\langle f_{k}\right\vert
\sum_{k^{\prime }}\left\vert g_{k^{\prime }}\right\rangle \left\langle
g_{k^{\prime }}\right\vert =1  \notag
\end{eqnarray}

and

\begin{eqnarray}
\mathcal{PT}\eta _{+}\mathcal{PT} &=&\mathcal{PT}\sum_{k}\left\vert
g_{k}\right\rangle \left\langle g_{k}\right\vert \mathcal{PT} \\
&=&\sum_{k}\left\vert g_{k}\right\rangle \left\langle g_{k}\right\vert =\eta
_{+}.  \notag
\end{eqnarray}%
On the other hand, equation (\ref{eta_elements}) allows the real matrix
representation of the metric operator by the transformation on the spatial
coordinate basis%
\begin{equation}
\left\vert l\right\rangle \rightarrow \left( \sqrt{-1}\right) ^{\text{mod}%
\left[ l,2\right] }\left\vert l\right\rangle .
\end{equation}%
For the sake of simplicity, hereafter $\eta _{+}$ denotes a real matrix.
Then from equations (\ref{eta_conjugate}) and (\ref{eta_PT}) indicate%
\begin{eqnarray}
\eta _{+}^{T} &=&\eta _{+} \\
\left( \eta _{+}\right) ^{-1} &=&R\eta _{+}R^{-1}, \\
\mathcal{P}\eta _{+}\mathcal{P} &=&\eta _{+},
\end{eqnarray}%
i.e., $\eta _{+}$ is a real unitary and bisymmetric\ matrix. Then the
eigenfunctions $\left\vert \epsilon _{n}\right\rangle $ of $\eta _{+}$\ can
always be written as real functions and obey%
\begin{equation}
\mathcal{P}\left\vert \epsilon _{n}\right\rangle =\pm \left\vert \epsilon
_{n}\right\rangle .
\end{equation}%
These properties is helpful for characterizing the equivalent physical
system of $H$.

Now we focus on the case of even $N$. A straightforward calculation shows
that eigenfunctions $\left\vert \epsilon _{n}\right\rangle $\ and $\mathcal{P%
}\left\vert \epsilon _{n}\right\rangle $\ have different parities. Then we
can reorder the canonical metric basis as%
\begin{equation}
\left\vert \epsilon _{n}\right\rangle =R\left\vert \epsilon
_{N+1-n}\right\rangle ,\text{ }n\in \left[ 1,N/2\right] ,  \label{R}
\end{equation}%
which satisfy%
\begin{eqnarray}
\mathcal{P}\left\vert \epsilon _{n}\right\rangle &=&\left\vert \epsilon
_{n}\right\rangle ,\text{ }n\in \left[ 1,N/2\right] , \\
\mathcal{P}\left\vert \epsilon _{n}\right\rangle &=&-\left\vert \epsilon
_{n}\right\rangle ,\text{ }n\in \left[ N/2+1,N/2\right] .  \notag
\end{eqnarray}%
Under such canonical metric basis, we have

\begin{eqnarray}
\mathcal{H}_{mn} &=&\sqrt{\frac{\epsilon _{m}}{\epsilon _{n}}}H_{mn} \\
&=&\sqrt{\frac{\epsilon _{m}}{\epsilon _{n}}}\left\langle \epsilon
_{m}\right\vert \mathcal{PT}H\mathcal{PT}\left\vert \epsilon
_{n}\right\rangle  \notag \\
&=&\sqrt{\frac{\epsilon _{m}}{\epsilon _{n}}}\left( -1\right) ^{\vartheta
(m)+\vartheta (n)}H_{mn}^{\ast }  \notag \\
&=&\left( -1\right) ^{\vartheta (m)+\vartheta (n)}\mathcal{H}_{mn}^{\ast },
\notag
\end{eqnarray}%
where
\begin{equation}
\vartheta (m)=\left\{
\begin{array}{c}
0\text{, }m\leq N/2 \\
1\text{, }m>N/2%
\end{array}%
\right. ,
\end{equation}%
is the Heaviside step function. On the other hand, since under the new
definition of basis $\left\vert l\right\rangle $ $H_{mn}$ is always
imaginary, all the elements $\mathcal{H}_{mn}$\ with $\vartheta
(m)+\vartheta (n)=0,$or $2$\ must vanish. Because $\left\vert \epsilon
_{n}\right\rangle $ can always be written as real, one can simply take $%
\left\vert \epsilon _{n}\right\rangle \rightarrow \left( \sqrt{-1}\right)
^{\vartheta (n)}\left\vert \epsilon _{n}\right\rangle $ to obtain the final
form of the equivalent Hermitian Hamiltonian
\begin{equation}
\mathcal{H}=\left[
\begin{array}{cc}
0 & A \\
A^{T} & 0%
\end{array}%
\right]
\end{equation}%
where $A$\ is real matrix and $A^{T}$\ is its transposed matrix.


\begin{thebibliography}{99}
\bibitem{remark} In this paper, we only give an exact proof for the even-$N$
system in the appendix. Numerical results imply that the conclusion
also holds for odd-$N$ case.
\end{thebibliography}
\end{document}